\documentclass[amsmath, amssymb, aps, 10pt, pra, twocolumn, floatfix, ]{revtex4}

\usepackage{graphicx}
\usepackage{dcolumn}
\usepackage{bm}
\usepackage{hyperref}
\usepackage{ulem}
\usepackage[dvipsnames]{xcolor}

\begin{document}
  \renewcommand{\vec}[1]{\mathbf{#1}}
  \newcommand{\Hssh}{H_{\text{SSH}}}
  \newcommand{\Hac}{H_{\text{ac}}}
  \newcommand{\hc}{\text{h.c.}}
  \newcommand{\pauli}{\widehat{\sigma}}
  \newcommand{\paulix}{\pauli_x}
  \newcommand{\pauliy}{\pauli_y}
  \newcommand{\pauliz}{\pauli_z}
  \newcommand{\paulii}{\pauli_0}
  \newcommand{\e}{\mathrm{e}}
  \newcommand{\abs}[1]{\left| {#1} \right|}
  \newcommand{\pump}{\text{p}}
  \newcommand{\omegap}{\omega_{\text{p}}}
  \newcommand{\omegash}{\omega_\text{s}}
  \newcommand{\Vol}{V_{\text{OL}}}
  \newcommand{\diff}{\mathrm{d}}
  \newcommand{\transpose}{\mathsf{T}}
  \newcommand{\Tmatrix}{\mathsf{T}}
  \newcommand{\Hdiag}{H_{\text{diag}}}
  \newcommand{\pt}{\widetilde{p}}
  \newcommand{\diag}[1]{\mathop{\mathrm{diag}}\nolimits\left( #1 \right)}

  \newcommand{\inset}[1]{\textcolor{black}{#1}}

	\preprint{APS/123-QED}

	\title{Resonant edge-site pumping of polaritonic Su-Schrieffer-Heeger lattices}

	\author{Yury Krivosenko} \email{y.krivosenko@gmail.com}

	\author{Ivan Iorsh}

  \author{Ivan Shelykh}

	\affiliation{ITMO University, St.\,Petersburg 197101, Russia}

	\date{\today}

	\begin{abstract}
    We have theoretically investigated Su-Schrieffer-Heeger chains modelled as optical lattices (OL) loaded with exciton-polaritons.
    The chains have been subject to the resonant pumping of the edge site and shaken in either adiabatic or high-frequency regime. The topological state has been controlled by the relative phases of the lasers constructing the OL.
    The dynamic problem of the occupation of the lattice sites and eigenstates has been semi-classically solved.
    Finally, the analysis of the occupation numbers evolution has revealed that gapless, topologically trivial and non-trivial chain configurations demonstrate perceptible behaviour from both qualitative (occupation pattern) and quantitative (total occupation) points of view.
	\end{abstract}

  \maketitle

  \section{Introduction}{\label{sec:Intro}}
    A topological insulator (TI) is a state of matter that behaves as a typical insulator in the bulk (with the band gap) but hosts midgap states localized at its boundary \cite{Hasan2010, Asboth2016} (the so-called \textsl{edge states}). These states are topologically protected in the sense that smooth varying system parameters cannot lead to a change of topological phase (trivial to non-trivial and vice versa) without closing the gap \cite{Franz2013}.

    The minimalistic model that manifests topological properties is one initially suggested by Su, Schrieffer, and Heeger (SSH) \cite{SSH1979}, which was invented to describe the conductivity in polyacetylene.
    The system can be presented as a dimerized linear chain each unit cell of which consists of two sites with staggered intra- ($t_1$) and inter-cell ($t_2$) hopping amplitudes.
    If the chain is finite and open boundary conditions are applied, the relation $\abs{t_1} > \abs{t_2}$ corresponds to the topologically trivial insulator (no edge states), $\abs{t_1} < \abs{t_2}$~-- to a TI, and $\abs{t_1} = \abs{t_2}$ implies the collapse of the band gap (see, e.g., \cite{Asboth2016, DalLago2015}).

    Due to the simplicity and rich properties of the model, its different variations have been extensively studied in recent years.
    Presented as a row of waveguides with on-site gain and loss, $i\gamma_{\text{A(B)}}$, a complicated SSH model was investigated with respect to topological protection of the midgap states and beam propagation along the guides \cite{Schomerus2013}. Also, Weimann et al. \cite{Weimann2017} considered topological protection of the bound states in the photonic SSH chain generalized by the presence of parity-time symmetric gain and loss that was achieved by putting alternating terms (${\pm} \, i \gamma/2$) onto the Hamiltonian diagonal.
    Li et al. \cite{Li2014} explored the SSH model extended by the next-nearest-neighbour hopping amplitudes. Marques and Dias \cite{Marques2017} researched multihole states in the fermionic chain with nearest-neighbour interactions.

    Periodic modulation of media is a recognized instrument for modifying the quantum lattice parameters, which is known as \textsl{Floquet engineering} and can give rise to a \textsl{Floquet TI} (both in electronics (fermionics) \cite{Lindner2011} and photonics \cite{Rechtsman2013}). Periodically forced, the system can be described by the effective Floquet Hamiltonian, $H_{\text{eff}} = H - i\hbar\partial_t$, that is independent of time and produces a quasienergy spectrum.

    Dynamics \cite{Dunlap1986, Bao1998, Liang2001} and quantum control \cite{Creffiled2007} of charged particles, as well as coherent destruction of tunnelling \cite{Li2015, VillasBoas2004, Luo2011} have been widely studied for periodically driven chains.
    G{\'o}mez-Le{\'o}n and Platero \cite{GomezLeon2013} scrutinized topology of the lattices driven by ac electric fields, obtained the general expressions for the renormalized system parameters and applied the results to the SSH chain. V. Dal Lago et al. \cite{DalLago2015} examined Floquet topological transitions and performed an analysis of the Zak phase applied to the SSH model. Hadad et al. \cite{Hadad2016} acquired the possibility of self-induced topological phase transitions in the chain with nonlinearities.
    Asb{\'o}th et al. \cite{Asboth2014} inspected the chiral symmetry of the periodically driven SSH model.

    In the field of cold atoms physics, a possible realization of Floquet engineering is utilization of cyclically modulated optical lattices \cite{Eckardt2017, Eckardt2005} (OL), particularly shaken ones \cite{Lignier2007, Hauke2012}.
    Another way to create an OL with a controllable topological state is to manage the relative phases or amplitudes of the lasers constructing the optical lattice \cite{Atala2013}.
    Stanescu et al. \cite{Stanescu2009} proposed a disc-shaped hexagonal OL with light-induced vector potential to realize the edge states, and to load boson into these states in order to probe the topology.

    Despite a notable progress in the domain of topological  insulators, little attention has been paid to the problem of pumped topological lattices. In this paper, we theoretically investigate the case of resonant pumping, i.\,e. when the frequency of the pumping field, $\omegap$, coincides with that of the cavity eigenmode ($\hbar\omegap = \varepsilon$). The SSH chain is brought about as an OL loaded with exciton-polaritons in microcavities \cite{Amo2010} and subject to shaking, the topology is adjusted by tuning the relative phase of the lasers constructing the optical potential. Both adiabatic and high-frequency shaking regimes are explored.

    The present paper is structured as follows. In Sec. \ref{sec:Theory}, we sketch the temporal problem of particle dynamics in the SSH lattice, introduce the formalism of pumping, and present the way of OL formation. Sec. \ref{sec:Discussion} is mostly concerned with the results of analytic and numeric simulations and consequent discussion. In Sec.~\ref{sec:Conclusion}, a brief conclusion is presented. Appendices \ref{sec:Appendix_analytic} and \ref{sec:Appendix_OL} contain the details on the dynamic problem solution and derivation of OL parameters.

  \section{Theory}{\label{sec:Theory}}
    Within the tight-binding approximation, behaviour of the SSH chain is described by the non-stationary Schr{\"o}dinger equation
    \begin{equation}{\label{eq:Schroedinger_nonstationary}}
      \Hssh \, | \psi(t) \rangle = i \, \partial_t | \psi (t) \rangle,
    \end{equation}
    where the SSH Hamiltonian in the basis of the lattice states is
    \begin{subequations}{\label{eq:Hssh_real}}
    \begin{align}
      \Hssh &= \sum\limits_{n} t_1 \left( | n, A \rangle \langle n, B | + \hc \right) \nonumber \\
      &+ \sum\limits_{n} t_2 \left( | n, B \rangle \langle n+1, A | + \hc \right)
      \label{eq:Hssh_nA,nB} \\
      &= \sum\limits_{n} t_1 | n \rangle \langle n | \otimes \paulix \nonumber \\
      &+ \sum\limits_{n} t_2 \left( | n+1 \rangle \langle n| \otimes \frac{\paulix + i \pauliy}{2} + \hc \right).
      \label{eq:Hssh_tensor_product}
    \end{align}
    \end{subequations}
    $| n, A \rangle$ and $| n, B \rangle$ in \eqref{eq:Hssh_nA,nB} denote the states of two sites in the $n$th unit cell. Later on, the external (cell-position, $| n \rangle$) and internal (in-cell, $| \alpha \rangle$) states are separated by means of a tensor product: $| n, \alpha \rangle \to |n\rangle \otimes | \alpha \rangle$. $\hbar = 1$ is accepted hereafter, $\sigma_i$ are Pauli matrices.

    To solve \eqref{eq:Schroedinger_nonstationary}, $\psi(t)$ is conventionally sought in the form
    \begin{equation}
      | \psi(t) \rangle = \sum\limits_{n} |n\rangle \otimes
      \begin{pmatrix}
        A_n (t) \\
        B_n (t)
      \end{pmatrix}
    \end{equation}
    (the explicit notation of the dependence on time is omitted hereafter).
    The SSH Hamiltonian acting on the state $| \psi \rangle$ yields
    \begin{align}
      \Hssh |\psi\rangle &= \sum\limits_{n}
      \left\{
      t_1 |n\rangle \otimes
      \begin{pmatrix}
        B_n \\
        A_n
      \end{pmatrix}
      \right. \nonumber \\
      &+ \left. t_2 \left[
      |n+1\rangle \otimes
      \begin{pmatrix}
        B_n \\
        0
      \end{pmatrix}
      + |n-1\rangle \otimes
      \begin{pmatrix}
        0 \\
        A_n
      \end{pmatrix}
      \right]
      \right\}.
    \end{align}
    Inserting this decomposition into \eqref{eq:Schroedinger_nonstationary}, we obtain the following set of equations
    \begin{equation}
      \label{eq:An_Bn_system}
      \left\{
        \begin{aligned}
          t_1 B_{n} + t_2 B_{n-1} &= i \, \partial_t A_n, \\
          t_1 A_n + t_2 A_{n+1} &= i \, \partial_t B_n, \quad n = 1\ldots N,
        \end{aligned}
      \right.
    \end{equation}
    which governs the behaviour of the system. Here, $N$ is the number of unit cells.

    \begin{figure}[b]
      \includegraphics[width=.9\columnwidth]{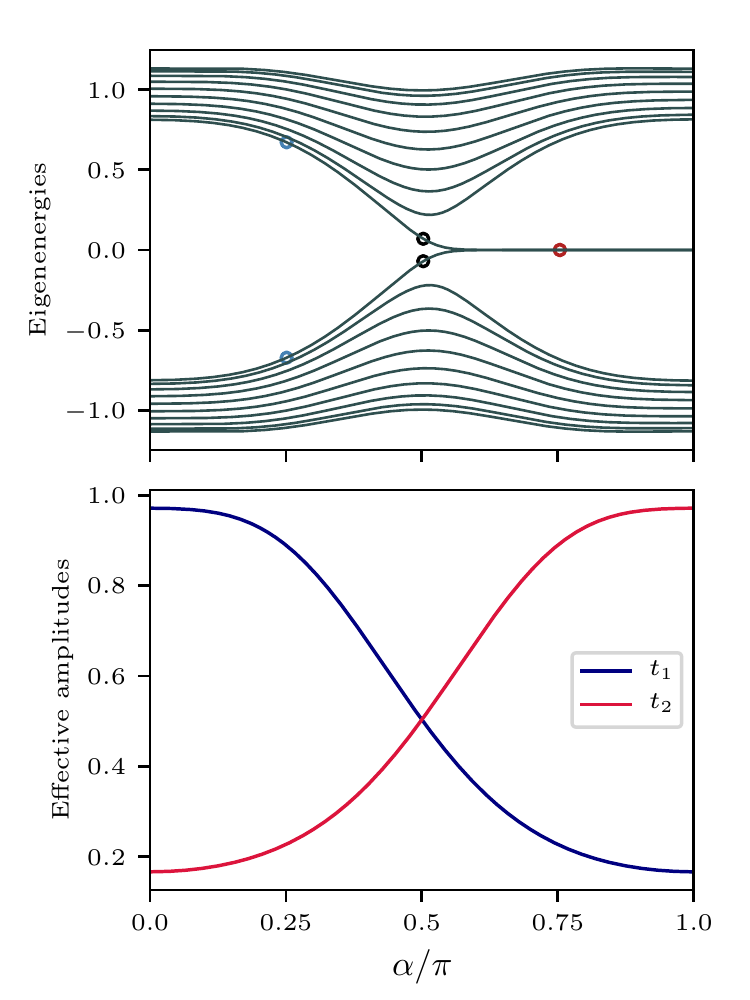}
      \caption{The band structures (upper panel) and effective amplitudes (lower panel) for the optical-lattice SSH chain. The dependences are plot versus the relative phase $\alpha$. The model parameters satisfy $V_0 \varkappa^2 = 2\mu = 0.5$. The number of the \inset{lattice sites equals $20$}. By the open circles in the upper panel, the states closest to the zero-energy for $\alpha = \pi/4, \pi/2, 3\pi/4$ are marked.}
      \label{fig:band_structures}
    \end{figure}

    Then, we move the focus of our consideration to the domain of non-interacting polaritons and phenomenologically add damping and pumping to the system as in \cite{Pervishko2013}. The former is introduced by means of the damping exponential factor $\gamma$ for all $n$ in \eqref{eq:An_Bn_system}, the latter -- by the pumping term $P \exp{i(-\omegap t + \phi_0)}$ placed into the equation relative to the pumped site $| n_0 \rangle {\otimes} | \text{C} \rangle$ (C equals A or B).
    These improvements alter eq.~\eqref{eq:An_Bn_system} to
    \begin{subequations}
      \label{eq:An_Bn_w/pump_damp}
      \begin{align}
        i \, \partial_t A_n &= t_1 B_{n} + t_2 B_{n-1} - i \, \gamma A_n + \varepsilon A_n
        \nonumber \\
        &+ \delta_{\text{A,C}} \, \delta_{n,n_0} \, P \exp{i(-\omegap t + \phi_0)}, \\
        i \, \partial_t B_n &= t_1 A_{n} + t_2 A_{n+1} - i \, \gamma B_n + \varepsilon B_n
        \nonumber \\
        &+ \delta_{\text{B,C}} \, \delta_{n,n_0} \, P \exp{i(-\omegap t + \phi_0)},
      \end{align}
    \end{subequations}
    where $P$, $\omegap$, and $\phi_0$ denote the amplitude, frequency, and relative phase of the pumping field, respectively, and $\varepsilon$, in the field of photonics, represents the cavity mode energy.
    Resonant pumping means the equality $\varepsilon = \omegap$ and is examined in Sec.~\ref{sec:Discussion}. The analytic solution of \eqref{eq:An_Bn_w/pump_damp} is presented in the Appendix~\ref{sec:Appendix_analytic} and is discussed in Sec.~\ref{sec:Discussion}.

    Further, we utilize the fact that the optical potential profile can be controlled by the relative phase of lasers forming the optical lattice. This provides tools to manage the topological phase transition in an optical lattice \cite{Atala2013}. Consider an OL formed by three laser fields, namely $E \e^{ikx+i\alpha}$, $E \e^{-ikx}$, and $E \e^{3ikx}$, where $\alpha$ is the relative phase of the first laser achieved by the frequency detuning ($\delta\omega$): $\alpha(\tau) = \delta\omega\, \tau$. Therefore, the optical potential produced by them is
    \begin{align}
      \Vol (x, \tau) &= V_0 \abs{\e^{ikx+i\alpha(\tau)} + \e^{-ikx} + \e^{3ikx}}^2 \nonumber \\
      &= V_0 \left( 3 + 4 \cos{\varkappa x} \, \cos{\alpha(\tau)} + 2 \cos{2\varkappa x} \right).
      \label{eq:Voptical}
    \end{align}
    Here, $\varkappa = 2k$, and $V_0$ is the dimensional constant. For this optical potential, the hopping amplitudes can be evaluated as
    \begin{align}
      t_i = t_i(\alpha) = \frac{\omega}{2} \cdot \e^{-\Delta_i^2} \cdot \left[ \Delta_i^2 + \frac12 \right],
      \label{eq:t_hop}
    \end{align}
    where
    \begin{subequations}{\label{eq:Delta's_t's(alpha)}}
      \begin{align}
        \Delta_1 (\alpha) &= \arccos{\left( \frac{\cos{\alpha}}{2} \right)}
        \biggl[ 2 V_0 \mu \varkappa^2 \left( 4 - \cos^2\alpha \right) \biggr]^{1/4},
        \label{eq:Delta_1(alpha)}
        \\
        \Delta_2 (\alpha) &= \Delta_1 (\pi - \alpha).
      \end{align}
    \end{subequations}
    Appendix~\ref{sec:Appendix_OL} contains the more detailed derivation.

  \section{Results and discussion}{\label{sec:Discussion}}
    \begin{figure*}[t]
      \includegraphics[width=.95\textwidth]{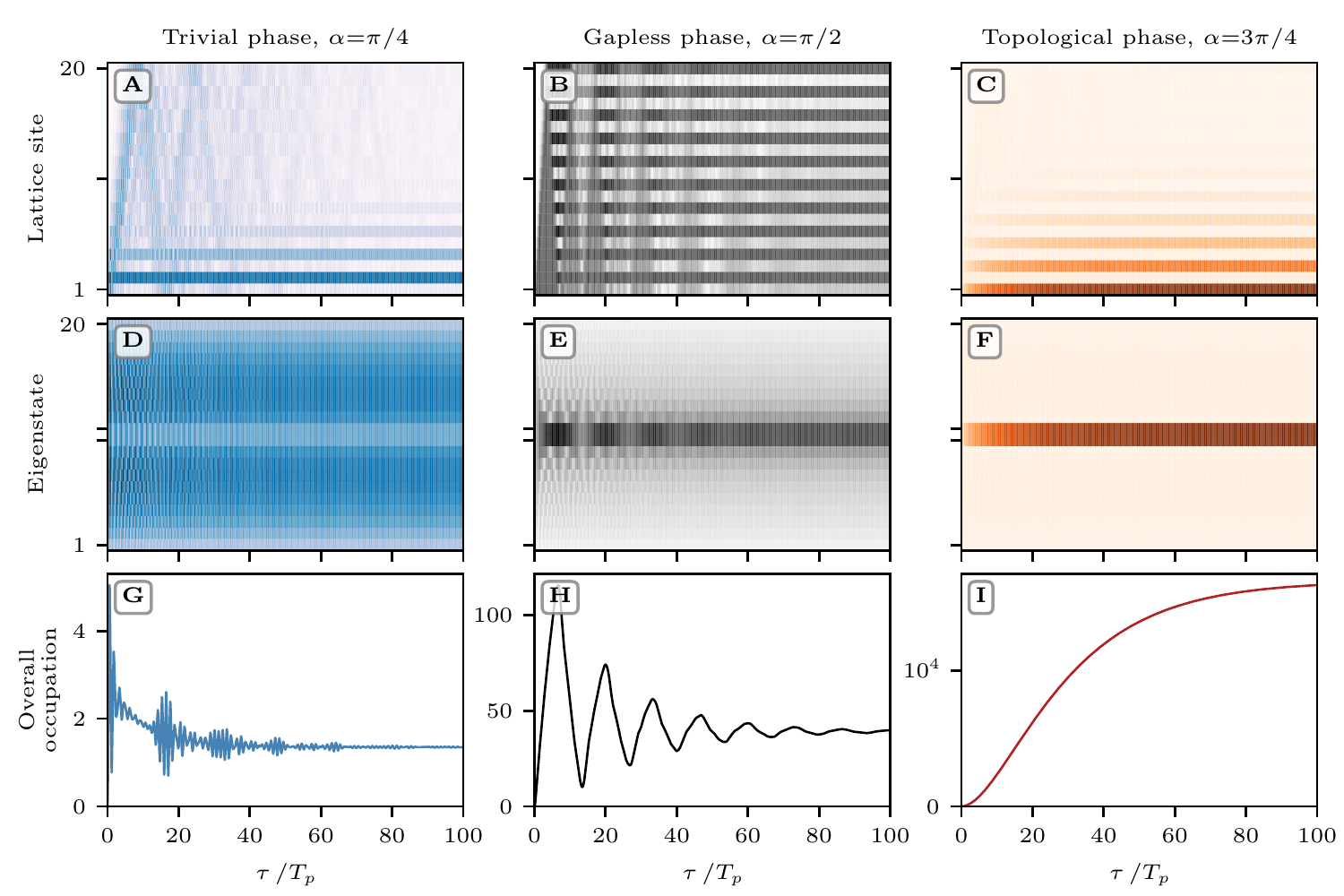}
      \caption{Dynamics of the occupation numbers of the topologically trivial (left column), gapless (middle column), and non-trivial (right column) SSH chains subject to resonant pumping of the terminal site.
      \inset{The magnitude of occupation numbers is displayed by means of color maps (subplots A--F: darker colours correspond to greater values, the colour normalization is power-law with the exponent $0.2$, which is chosen for better readability).
      The top row (A--C): evolution of the occupation numbers of the lattice sites (lattice site 1 is the pumped one).
      The middle row (D--F): evolution of the occupation numbers of the eigenstates.
      The bottom row (G--I): evolution of the total occupation of the chains.
      The tick labels in the middle of ordinate axes in subplots A--F indicate the centre of the chain (A--C) and the pairs of sites closest to the zero energy (D--F) with the latest being marked by the open circles in Fig.~\ref{fig:band_structures}. In all the subplots, time is measured in pumping cycles, $T_{\text{p}}$.}
      The additional, relative to Fig.~\ref{fig:band_structures}, model parameters are $\lambda = 0.0075$, $\varepsilon = \omegap = 1$.}
      \label{fig:adiabatic}
    \end{figure*}
    Fig.~\ref{fig:band_structures} shows typical dependences of the effective hopping amplitudes (the bottom subplot) and corresponding band structure (the top subplot) obtained by the numerical simulation. The dependences are plotted versus the relative phase $\alpha$. The band structures are computed as solutions of the stationary Schr{\"o}dinger equation with Hamiltonian $\Hssh$ \eqref{eq:Hssh_tensor_product} where the amplitudes $t_1$ and $t_2$ are substituted by the effective values \eqref{eq:t_hop}. The parameters taken for the simulation are $V_0 \varkappa^2 = 2\mu = 0.5$,
    the number of unit cells equals $10$ \inset{thus providing $20$ lattice sites}.
    We can see that the whole area is split into two parts, $\alpha {\in} [0, \pi/2)$ and $\alpha {\in} (\pi/2, \pi]$, which relate to topologically trivial and non-trivial configurations, respectively. $\alpha {=} \pi/2$ indicates no-band-gap phase.

    As can be inferred from eqs. \eqref{eq:t_hop} and \eqref{eq:Delta's_t's(alpha)} and seen in Fig.~\ref{fig:band_structures}, the hopping amplitudes do not reach zero at finite non-zero values of parameters $V_0$, $\varkappa$, and $\mu$ within the chosen pattern of the lattice formation. Thus, the coherent destruction of tunnelling is not feasible here. On the other hand, $V_0 \to 0$ or $V_0 \to + \infty$ definitely leads to zero-value hopping amplitudes but, firstly, any of these conditions sets both $t_1$ and $t_2$ to zero simultaneously and for all $\alpha$ and, secondly, these conditions would signify no optical lattice at all or a lattice with infinite potential barriers, respectively.

    In the remainder of this Section, we scrutinize the problem of pumping the terminal site ($|1, \text{A} \rangle$) of the chain subject either to adiabatic ($\delta\omega {\ll} t_1, t_2$) or high-frequency ($\delta\omega {\gg} t_1, t_2$) shaking.

    Fig.~\ref{fig:adiabatic} represents the solution of the adiabatic dynamic problem \eqref{eq:An_Bn_w/pump_damp} for $\alpha = \pi/4$ (left panels), $\pi/2$ (middle panels), and $3\pi/4$ (right panels) at the zero initial conditions. The hopping amplitudes are $t_1 = 0.89$ and $t_2 = 0.23$ in the trivial case and vice versa in the non-trivial one. Both $t_1$ and $t_2$ equal $0.50$ for the gapless mode.
    The solution is semi-analytic in the sense that the problem of diagonalization of $\Hssh$ is numerically solved, whereas the evolution of occupation numbers is then analytically calculated (see Appendix~\ref{sec:Appendix_analytic}). The detailed description of the graphs is given in the figure caption.
    We notice here that the condition of localization \eqref{eq:App_r0<<lambda} can be met particularly at $V_0 = 1/2$, $\mu = 1/4$, and $\varkappa = 1$.

    \begin{figure}[b]
      \includegraphics[width=.95\columnwidth]{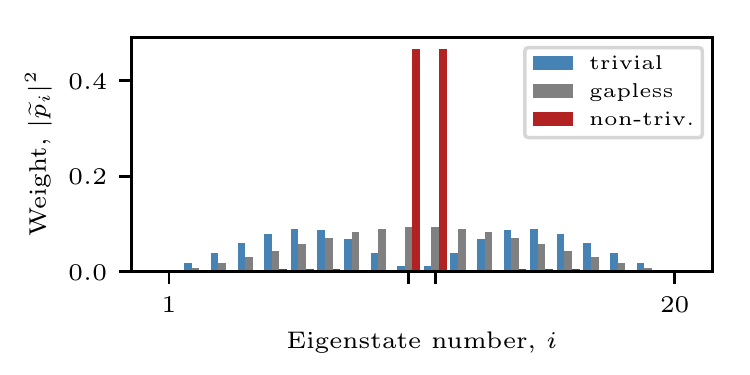}
      \caption{The weights, $\abs{\pt_i}^2$, of decomposition of the $| 1, A \rangle$ state over the eigenstates, $i$, for the trivial (blue bars), \inset{gapless (grey bars)}, and non-trivial (red bars) chains.}
      \label{fig:bar_plot}
    \end{figure}
    From the data in this figure, we can conclude that the evolution of occupation numbers differs both qualitatively and quantitatively for the trivial and non-trivial phases. These differences mainly arise from the presence (or absence) of the edge states susceptible for such pumping and can be explained in terms of eigenenergies, eigenstates and equations \eqref{eq:App_|y_i|^2} and \eqref{eq:App_|y_i|^2_t->infty}. For the edges states localized well enough, $\varepsilon_i \simeq 0$ and the values $\abs{\pt_i}^2$ remarkably exceed all others (as the terminal site $| 1, \text{A} \rangle$ being decomposed over the eigenstates mostly consists of the edge states wavefunctions). This retains almost only $\gamma^2$ in the corresponding denominators in \eqref{eq:App_|y_i|^2} and \eqref{eq:App_|y_i|^2_t->infty}, which causes considerable growth of these terms. These reasonings are supported by Figures \ref{fig:adiabatic}D, \ref{fig:adiabatic}E, \ref{fig:adiabatic}F, and \ref{fig:bar_plot}. Fig.~\ref{fig:adiabatic} D, E, and F show how the occupations of eigenstates evolve. As one can see, almost only the edge states are pumped in the non-trivial case (\ref{fig:adiabatic}F) and they are pumped much more intensively compared to the whole trivial (\ref{fig:adiabatic}D) and gapless (\ref{fig:adiabatic}E) chains.
    Fig.~\ref{fig:bar_plot} demonstrates the weights $\abs{\pt_i}^2$ for the trivial (blue bars), gapless (gray bars), and non-trivial (red bars) chain configurations
    \inset{The abscissa axis herein (eigenstate number, $i$) totally coincides with the ordinate axis of Fig.~\ref{fig:adiabatic}, panels D--F.}
    In the topological case, the edge states clearly dominate the others (the two highest red vertical bars, $\abs{\pt}^2 \simeq 0.47$ for each), whereas the distribution is more uniform and does not possess such pronounced singularities in the both trivial and gapless cases (the values $\abs{\pt}^2$ do not exceed $0.1$). As a result, the final summary occupation \eqref{eq:App_Final_sum_occupation} of the non-trivial chain exceeds those of trivial and gapless ones ca. $12{\cdot}10^3$ and $400$ times, respectively (cf. Figures \ref{fig:adiabatic}G, \ref{fig:adiabatic}H, and \ref{fig:adiabatic}I).

    The gapless case exhibits an individual pattern relative to both trivial and non-trivial phases (Fig.~\ref{fig:adiabatic}B): since a certain moment, the lattice sites become occupied in a staggered way and this distribution remains nearly uniform along the chain. What is also interesting in the pattern is that the pumped site finally stays almost unfilled despite the fact it is kept on pumped (as in the trivial configuration). As regards to the eigenstates of the gapless phase, they are populated similarly to the topological phase, cf. Fig.~\ref{fig:adiabatic}E and \ref{fig:adiabatic}F. But the difference consists in the fact that there are no edge states in the gapless phase and the two mid-energy states (marked by the pair of horizontal tick labels in Fig.~\ref{fig:adiabatic}E and by the black open circles in Fig.~\ref{fig:band_structures}) are just the delocalized ones which are the closest to the zero energy.

    \begin{figure}[b]
      \includegraphics[width=.95\columnwidth]{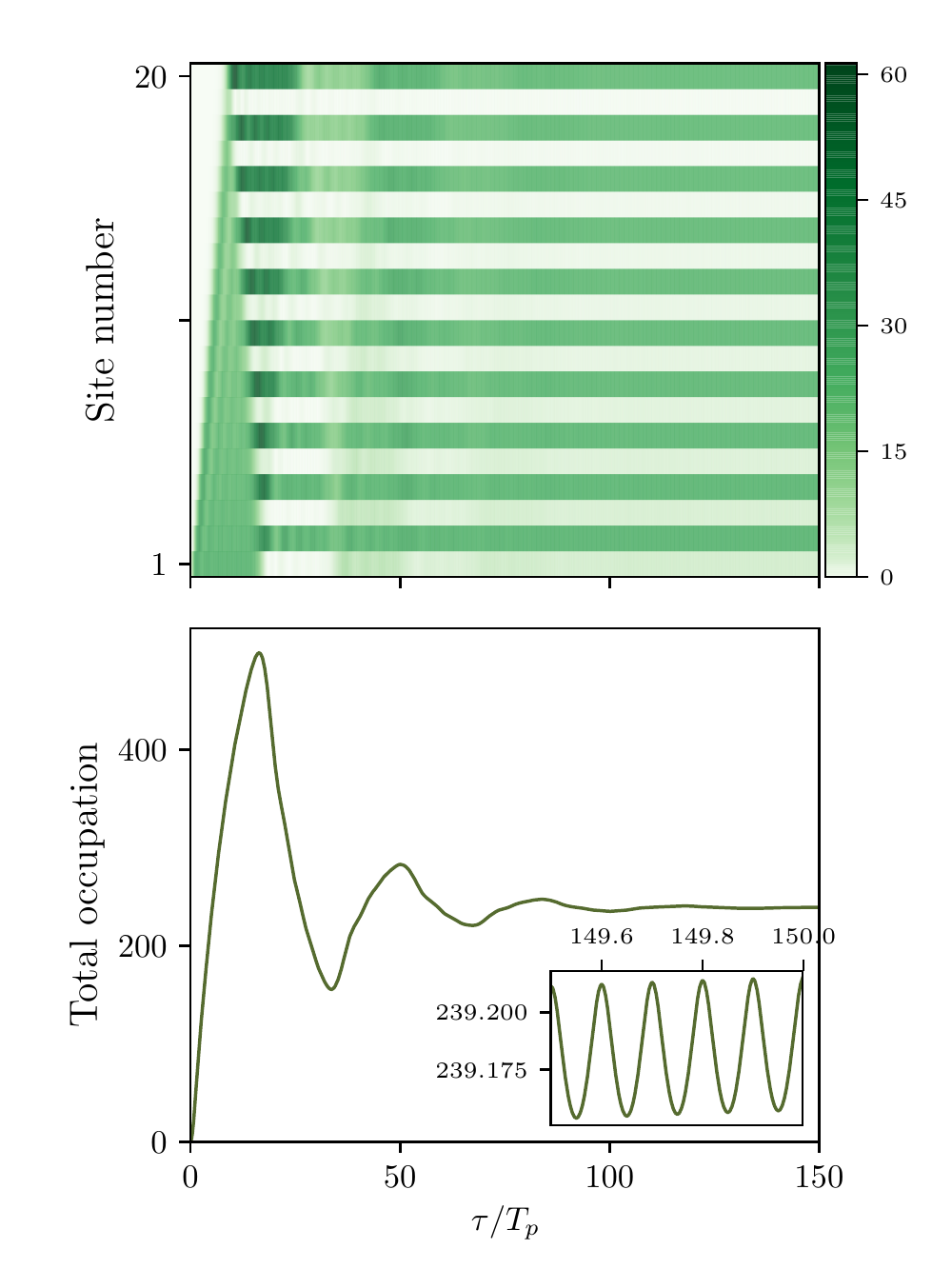}
      \caption{Evolution of the sites occupation within the high-frequency regime. Upper panel: the colour-map representation of the each site occupation versus time. Darker colors correspond to greater values. Lower panel: the total occupation of the chain. The inset represents the zoomed segment for the last pumping cycle half within the examined time span. In both panels, time is measured in the pumping cycles, $T_\text{p}$.}
      \label{fig:high_freq_dynamic}
    \end{figure}

    In the end, we examine the case of high-frequency shaking, see Fig.~\ref{fig:high_freq_dynamic}. For the numeric calculation, the shaking frequency has been chosen as $\omegash = 10 \, \omegap = 10$. All other parameters have remained unchanged. The data evidence a qualitative similarity between the gapless adiabatic (Fig.~\ref{fig:adiabatic} B and H) and high-frequency (Fig.~\ref {fig:high_freq_dynamic}) regimes. This resemblance can be interpreted in the way rather inherent to problems of high-frequency Floquet engineering:
    \inset{
    such periodic modulation of Hamiltonian is known to effectively renormalize the hopping amplitudes in arrays.
    For example, the hopping amplitudes of the SSH chain driven by a high-frequency ac field are modified by multiplying by the correpsoding Bessel functions, see \cite{GomezLeon2013}.
    In our case, parameters $\Delta_{1(2)}$ \eqref{eq:Delta's_t's(alpha)} and therefore amplitudes $t_{1(2)}$ \eqref{eq:t_hop} follow the periodicity of the relative phase $\alpha(\tau)$ thus making the Hamiltonian $2\pi/\delta\omega$-periodic and Floquet theory applicable. Hence,
    we suppose that it is possible to get an appropriate renormalization of $t_1$ and $t_2$ remaining them equal and giving rise to a better agreement between the patterns.
    Of course, such renormalization would not reproduce the high-frequency oscillations as those displayed in the inset in the lower panel  of Fig.~\ref{fig:high_freq_dynamic}.
    }

  \section{Conclusion}{\label{sec:Conclusion}}
    The main aim of the present study was to theoretically examine the response of topologically different Su-Schrieffer-Heeger chains to resonant pumping of the edge site.
    The chains were modelled as the condensate of non-interacting exciton-polaritons in the shaken optical lattice. The chain topological phase was managed by the relative phases of the lasers constructing the lattice.
    The intersite hopping amplitudes were analytically calculated within the harmonic approximation to the local site potential. For the adiabatic regime of shaking, the dynamic problem has been semi-analytically solved.

    Finally, the analysis of the occupation numbers evolution has shown that, under the considered conditions, the gapless, topologically trivial and non-trivial chains exhibit perceptible behaviour from both qualitative and quantitative points of view.
    Firstly, they demonstrate distinct patterns of the sites occupation and, secondly, their susceptibilities to the pumping are significantly different.

    We believe that our findings could be useful for the purposes of detecting the topological phase of matter also in more general systems than the minimalistic one considered here in detail.


  \begin{appendix}
    \section{Analytic solution of the pumping problem}{\label{sec:Appendix_analytic}}
      Consider the dynamic problem \eqref{eq:An_Bn_w/pump_damp} in more detail. Denoting the vector $\left( A_1, B_1, A_2, B_2, \ldots A_N, B_N \right)^{\transpose}$ by $x$, we rewrite the system of equations in the matrix form:
      \begin{gather}
        i \, \dot{x} = \Hssh \, x + (\varepsilon - i\gamma) \, x + p \, \exp{i (-\omegap t)},
        \label{eq:App_Hssh_dynamic_x}
      \end{gather}
      where $p$ is the vector that represents pumping. For the pumping of the edge site, $p = P \, \e^{i\phi_0} \, (1, 0, \ldots 0)$. Suppose $\Tmatrix$ is the diagonalization matrix of $\Hssh$, so that $\Tmatrix^\dagger \Hssh \Tmatrix = \diag{\varepsilon_1, \varepsilon_2, \ldots \varepsilon_N} = \Hdiag$. After the transition to the eigenstates $y$ by means of the substitution
      \begin{equation}{\label{eq:App_x=Ty}}
        x = \Tmatrix y
      \end{equation}
      and left multiplying by $\Tmatrix^\dagger$, one switches to the set of equations
      \begin{gather}
        i \, \dot{y} = \Hdiag \, y + (\varepsilon - i\gamma) \, y + \Tmatrix^\dagger p \, \e^{-i\omegap t},
        \intertext{which can be then separated into $N$ independent ones}
        i \, \dot{y}_i = (\varepsilon_i + \varepsilon - i\gamma) \, y_i + \pt_i \, \e^{-i \omegap t}
        \label{eq:App_evo_y_sep}
      \end{gather}
      for $i = 1\ldots 2N$, where $\pt = \Tmatrix^\dagger p$. The solution of each of \eqref{eq:App_evo_y_sep} is sought as the sum of solutions of the corresponding homogeneous and inhomogeneous equations. Applying the zero initial conditions, we finally arrive at
      \begin{equation}
        y_i (t) = \frac{\pt_i}{\varepsilon_i + \varepsilon - \omegap - i\gamma} \cdot \left[ \e^{-i\omegap t} - \e^{-i(\varepsilon_i + \varepsilon)t - \gamma t} \right].
        \label{eq:App_y_final}
      \end{equation}
      The latter equation enables us to easily get the limit value of the $i$th mode norm:
      \begin{align}
        \abs{y_i(t)}^2 &= \frac{\abs{\pt_i}^2}{(\varepsilon_i + \varepsilon - \omegap)^2 + \gamma^2} \times
        \nonumber \\
        &\times \left[ 1 + \e^{-2\gamma t} - 2 \, \e^{-\gamma t} \cos{(\varepsilon_i + \varepsilon - \omegap)t} \right]
        \label{eq:App_|y_i|^2}
        \\
        &\xrightarrow[t \to +\infty]{} \frac{\abs{\pt_i}^2}{(\varepsilon_i + \varepsilon - \omegap)^2 + \gamma^2}.
        \label{eq:App_|y_i|^2_t->infty}
      \end{align}
      Thus, the overall occupation of the chain at long time scales is
      \begin{align}
        \sum\limits_{n = 1}^{2N} &\abs{x_n(t)}^2 =
        \sum\limits_{n = 1}^{2N} \abs{y_n(t)}^2 \xrightarrow[t \to +\infty]{}
        \nonumber \\
        &\to
        \sum\limits_{n = 1}^{2N} \frac{\abs{\pt_i}^2}{(\varepsilon_i + \varepsilon - \omegap)^2 + \gamma^2}.
        \label{eq:App_Final_sum_occupation}
      \end{align}
      The real space amplitudes, $x$, at each site are then calculated in compliance with \eqref{eq:App_x=Ty}.

      For further comparison, we solve \eqref{eq:App_Hssh_dynamic_x} and \eqref{eq:App_evo_y_sep} at the absence of pumping but with the specific initial conditions:
      $x(t{=}0) = (1, 0\ldots 0)^\transpose$,
      which means the initial occupation of the terminal site. Solution of this dynamic problem can be reduced to
      \begin{equation}{\label{eq:App_y(t)_no_pumping}}
        y_i (t) = \frac{\pt_i}{\abs{\pt}} \cdot \e^{-i(\varepsilon_i + \varepsilon)t - \gamma t}.
      \end{equation}

    \section{Hopping amplitudes}{\label{sec:Appendix_OL}}
      For the optical potential \eqref{eq:Voptical}, one can express the positions of minima as
      \begin{equation}{\label{eq:App_Voptical_minima}}
        x_n^{\pm}(\alpha) = \frac{2\pi}{\varkappa} \left( n + \frac12 \pm \frac{\arccos{\frac{\cos{\alpha}}{2}}}{2\pi} \right),
      \end{equation}
      which are attributed to the wells that host the lattice sites. The intracell, $\Delta x_1$, and intercell, $\Delta x_2$, distances between the neighbour sites are:
      \begin{subequations}{\label{eq:App_dx_both}}
        \begin{align}
          \Delta x_1(\alpha) &= x_n^+ - x_n^- = \frac{2 \, \arccos{\frac{\cos{\alpha}}{2}}}{\varkappa} \label{eq:App_dx1} \\
          \Delta x_2(\alpha) &= \frac{2\pi}{\varkappa} - \Delta x_1(\alpha). \label{eq:App_dx2}
        \end{align}
      \end{subequations}

      In order to evaluate the hopping amplitudes $t_1 = t(x_n^- {\to} x_n^+)$ and $t_2 = t(x_n^+ {\to} x_{n+1}^-)$, we use the harmonic approximation for the optical potential \eqref{eq:Voptical} in the vicinity of the minima \eqref{eq:App_Voptical_minima}. That leads to the magnitudes of the vibration frequency, $\omega$, and the zero-vibrations amplitude, $r_0$:
      \begin{subequations}{\label{eq:App_omega&r0}}
        \begin{align}
          \omega(\alpha) &= \sqrt{\frac{1}{\mu} \cdot \frac{\partial^2 \Vol}{\partial x^2}} = \sqrt{\frac{2 V_0 \varkappa^2 \left( 4 - \cos^2{\alpha} \right)}{\mu}},
          \label{eq:App_omega_as_derivative}
          \\
          r_0(\alpha) &= \sqrt{\frac{1}{\mu\omega(\alpha)}} = \biggl[ 2 V_0 \mu \varkappa^2 \left( 4 - \cos^2\alpha \right) \biggr]^{-1/4},
          \label{eq:App_r0(alpha)}
        \end{align}
      \end{subequations}
      where $\mu$ is the oscillator reduced mass. The vibrations frequency $\omega$ remains the same for all the minima.
      The localized Wannier states are taken as the zero vibrational level harmonic wavefunctions
      \begin{equation}{\label{eq:App_Chi_0}}
        \chi_0 (x) = \frac{1}{r_0^{1/2} \, \pi^{1/4}} \, \e^{-x^2/2r_0^2}
      \end{equation}
      centred at the corresponding minima. In this case, the hopping amplitudes are calculated as
      \begin{align}
        t_i = t(\Delta x_i) &= \int\limits_{-\infty}^{+\infty} \diff x \ \chi_0 (x+\Delta x_i) \, \frac{\mu\omega^2 \, x^2}{2} \, \chi_0 (x) \nonumber \\
        &= \frac{\omega}{2} \cdot \e^{-\Delta_i^2} \cdot \left[ \Delta_i^2 + \frac12 \right],
        \label{eq:App_t_hop}
      \end{align}
      where $\Delta_i = \Delta_i(\alpha) = \Delta x_i(\alpha)/2r_0(\alpha)$ (see \eqref{eq:App_dx_both}, \eqref{eq:App_r0(alpha)}), and $\omega$ is regarded a function of $\alpha$ as well \eqref{eq:App_omega_as_derivative}.

      The dependences of $\Delta x_{1(2)}$ \eqref{eq:App_dx_both} and $\omega$ \eqref{eq:App_omega&r0} on $\alpha$ disclose that
      \begin{subequations}{\label{eq:App_omega_dx_alpha}}
        \begin{align}
          \omega (\pi \pm \alpha) &= \omega (\alpha),
          \quad
          r_0 (\pi \pm \alpha) = r_0 (\alpha),
          \label{eq:App_omega(alpha)}
          \\
          \Delta x_1 (\pi \pm \alpha) &= \Delta x_2 (\alpha),
          \quad
          \Delta x_2 (\pi \pm \alpha) = \Delta x_1 (\alpha). \label{eq:App_dx(alpha)}
        \end{align}
      \end{subequations}
      Therefore,
      \begin{subequations}{\label{eq:App_Delta's_t's(alpha)}}
        \begin{align}
          \Delta_1 (\alpha) &= \frac{\Delta x_1(\alpha)}{2r_0(\alpha)} =
          \arccos{\left( \frac{\cos{\alpha}}{2} \right)} \nonumber \\
          & \times
          \biggl[ 2 V_0 \mu \varkappa^2 \left( 4 - \cos^2\alpha \right) \biggr]^{1/4},
          \label{eq:App_Delta_1(alpha)}
          \\
          \Delta_2 (\alpha) &= \Delta_1 (\pi - \alpha),
          \intertext{and}
          t_2 (\alpha) &= t_1 (\pi - \alpha). \label{eq:App_t1_t2(alpha)}
        \end{align}
      \end{subequations}

      The condition of the on-site localization of the Wannier states, $r_0 \ll 2\pi/\varkappa$, would require
      \begin{equation}{\label{eq:App_r0<<lambda}}
        \frac{\varkappa^2}{2 V_0 \mu \left( 4 - \cos^2{\alpha} \right) (2\pi)^4} \ll 1
      \end{equation}
      for all the phases $\alpha$ within the operation region.
  \end{appendix}

\end{document}